\documentclass[sn-mathphys,Numbered]{sn-jnl} 

\usepackage{subcaption}
\usepackage{graphicx}
\usepackage{multirow}
\usepackage{booktabs}
\usepackage{array}
\usepackage{amsmath,amssymb,amsfonts}
\usepackage{amsthm}
\usepackage{mathrsfs}
\usepackage[title]{appendix}
\usepackage{xcolor}
\usepackage{textcomp}
\usepackage{manyfoot}
\usepackage{algorithm}
\usepackage{algpseudocode}
\usepackage{listings}
\usepackage{hhline}
\usepackage{multicol}
\usepackage{enumerate}
\usepackage{natbib}
\usepackage{float}

\PassOptionsToPackage{hyphens}{url} 
\usepackage{hyperref} 

\DeclareUnicodeCharacter{202F}{\,}

\theoremstyle{thmstyleone}

\theoremstyle{thmstyletwo}

\theoremstyle{thmstylethree}

\begin{document}
	
	\title[]{
		Destroying the Kerr Newman MOG
		Black hole with Scalar Test Field}
	
	
	\author[]{\fnm{Waqar} \sur{Ahmad}}\email{dmt203001@cust.pk}
	\author[]{\fnm{Abdul Rehman} \sur{Kashif}}\email{dr.arehman@cust.edu.pk}
	\author[]{\fnm{Ayyesha} \sur{K.Ahmed}}\email{ayyesha.kanwal@cust.edu.pk}

	\affil[]{\orgdiv{Department of Mathematics}, \orgname{Capital University of Science and Technology}, \city{Islamabad}, \country{Pakistan}}
	



\abstract{We test the weak cosmic censorship conjecture (WCCC) for the Kerr–Newman–modified gravity black hole (KN--MOG--BH) by interacting scalar test fields with the KN–MOG–BH. Neglecting backreaction effects, we first show that the scalar test fields with frequencies just above the superradiance threshold can overspin both extremal and nearly extremal KN–MOG–BHs, leading to the formation of naked singularities. Moreover, test fields can continuously push nearly extremal KN--MOG--BHs toward their extremal limit. Next, we incorporate backreaction effects, assuming that the event horizon’s angular velocity rises prior to the test field’s absorption. We show that backreaction prevents overspinning in the extremal KN--MOG--BH, whereas in the nearly extremal case, it fails to stop the BH from being overspun.}

\maketitle

\section{Introduction}\label{intro}

The cosmic censorship conjecture suggests that spacetime singularities formed by gravitational collapse are always concealed within an event horizon \cite{hawking1970singularities}. The cosmic censorship conjecture, asserting that naked singularities cannot exist in our universe, is stated in two versions: weak and strong
\cite{joshi2007gravitational}. The WCCC proposes that the end state of gravitational collapse is a BH rather than a naked singularity \cite{penrose1969collapse,christodoulou1999global}. This means that singularities formed in gravitational collapse must always be hidden behind an event horizon, preventing their visibility to distant observers. The strong cosmic censorship (SCCC) extends this requirement by forbidding both naked and locally naked singularities, ensuring that no observer—whether at infinity or within a black hole—can access such regions \cite{penrose1979singularities}. For more than fifty years, the conjecture has remained unproven, yet it can still be tested. There are
different methods to check the validity of the WCCC. One way of testing the WCCC is to add a scalar field or to throw a test particle into an existing black hole. If the scalar field or test particle passes through the horizon and changes the parameters of the black hole such that the horizon disappears, then the WCCC might break down. If the conjecture is violated
somehow, the naked singularity may open a door to test the theories of quantum gravity.

The conjecture was first examined through a thought experiment introduced by Wald \cite{wald1974gedanken}, who demonstrated in a study that a test particle cannot destroy the horizon of an extremal Kerr–Newman black hole. Since then, numerous researchers have extended this line of investigation, employing Wald’s gedanken experiment framework to reassess possible violations of the WCCC. Several of these test-particle based analyses have reported scenarios leading to the formation of naked singularities \cite{jacobson2009overspinning, hubeny1999overcharging,richartz2008overspinning, matsas2009can,richartz2011challenging}. This conclusion was later generalized to scalar test fields \cite{semiz2011dyonic, toth2012test}.

 Hubeny \cite{hubeny1999overcharging} later introduced an alternative approach in which one begins analysis with a nearly extremal black hole rather than an  extremal one. She demonstrated that, by employing suitably chosen charged particles, it may be possible to overcharge a nearly extremal Reissner–Nordström black hole. This line of reasoning was subsequently extended to Kerr and Kerr–Newman black holes \cite{saa2011destroying}. To address these scenarios, backreaction effects were later taken into account, showing that such effects can prevent the destruction of the horizon \cite{barausse2010test,isoyama2011cosmic,zimmerman2013self}.

The potential violation of the WCCC through quantum tunneling of test particles has been examined in several works \cite{matsas2007overspinning,hod2008weak,hod2008return}. In addition, the stability of event horizons in asymptotically anti-de Sitter spacetimes has been analyzed by introducing perturbations with test particles and fields \cite{gwak2016cosmic,chen2019thermodynamics}.

Recent studies have also explored possible violations of the WCCC through scalar field perturbations. In the absence of backreaction effects, bounds on the mode frequencies were derived under the assumption that the scalar field is absorbed by the black hole, potentially leading to the destruction of the horizon \cite{bekenstein1973extraction,hong2019thermodynamics,duztacs2020testing,gwak2018weak,duztacs2019kerr}. Based on these frequency ranges, it was argued that an extremal Kerr–Newman black hole could approximately violate the WCCC.

In this paper, we test the WCCC in KN–MOG–BH using a scalar field probe. In Section~(\ref{sec:1}), we introduce the KN–MOG–BH metric and analyze overspinning in the extremal and nearly extremal cases without considering backreaction effects. In Section~(\ref{effects}), we incorporate backreaction effects to examine the validity of the WCCC for extremal and nearly extremal cases. Finally, in Section~(\ref{conclusion}), we present our conclusions.
\section{Kerr-Newman-MOG black hole and scalar field dynamics}
\label{sec:1}


In Boyer--Lindquist coordinates, the KN--MOG line element can be expressed as \cite{moffat2015kerr}
\begin{equation}
	\label{metric_modified}
	ds^{2} = -\left[\,dt - a\sin^{2}\theta\,d\varphi\right]^{2}\frac{\triangle}{\rho^{2}}
	+\frac{\rho^{2}}{\triangle}\,dr^{2} 
	+\rho^{2}\,d\theta^{2} 
	+\frac{\sin^{2}\theta}{\rho^{2}}\left[(r^{2}+a^{2})\,d\varphi - a\,dt\right]^{2}
\end{equation}

where
\begin{equation}
	\label{rhoDelta_modified}
	\triangle = r^{2} + a^{2} + Q^{2} - 2M(1+\alpha)rG_{N} + \alpha M^{2}(1+\alpha)G_{N}^{2},
	\qquad
	\rho^{2} = r^{2} + a^2\cos^{2}\theta.
\end{equation}

The dimensionless parameter $\alpha$ quantifies the deviation of the modified gravity constant $G$ from the Newtonian constant $G_{N}$,
\begin{equation}
	\label{alpha}
	\alpha = \frac{G - G_{N}}{G_{N}}.
\end{equation}

The angular momentum of KN--MOG--BH is related to the Arnowitt-Deser-Misner (ADM) mass \cite{sheoran2018mass}
\begin{equation}
	\label{ADM}
	\mathcal{M} = (1+\alpha)\,M.
\end{equation}

It is convenient to rewrite $\triangle$ in terms of $\mathcal{M}$. Setting $G_{N}=1$ (geometric units), one finds
\begin{equation}
	\label{DeltaM}
	\triangle = r^{2} - 2\mathcal{M}\,r + a^{2} + Q^{2} + \frac{\alpha}{1+\alpha}\,\mathcal{M}^{2}.
\end{equation}

The event horizon radii correspond to the real roots of $\triangle=0$, given by
\begin{equation}
	\label{rH}
	r_{H} = \mathcal{M} \pm \sqrt{\frac{\mathcal{M}^{2}}{1+\alpha} - a^{2} - Q^{2}}.
\end{equation}

Thus, the spacetime represents a black hole only if the parameters satisfy the condition
\begin{equation}
	\label{condition}
	\mathcal{M}^{2} \;\geq\; (1+\alpha)\left(a^{2}+Q^{2}\right),
\end{equation}
with equality corresponding to the extremal case, while violation of this inequality would result in a naked singularity.

A scalar test field scatters off the KN--MOG--BH as described by \cite{wondrak2018superradiance}
\begin{equation}
	\Phi(t, r, \theta, \phi) 
	= R(r)\, Z(\theta, \phi)\, e^{-i\omega t},
\end{equation}
where $\omega$ is the frequency of the scattering field, 
$R(r)$ is the radial function, and $Z(\theta, \phi)$ represents oblate spheroidal harmonic 
which can be expressed in terms of the oblate spheroidal wave function $S(\theta)$ 
\begin{equation}
	Z(\theta, \phi) = S(\theta)\, e^{i m \phi}.
\end{equation}
We consider a charged scalar field propagating in the KN–MOG spacetime. Each mode of the field is characterized by a frequency $\omega$ and an azimuthal quantum number $m = 0, \pm 1, \pm 2, \ldots$. The radial and angular dependence of the field is assumed such that it behaves an ingoing wave near the black hole horizon, whereas it has both ingoing and outgoing
components far from it. Upon calculating the stress–energy tensor, one obtains
\begin{equation}\label{1}
	\frac{\triangle J}{\triangle E} = \frac{m}{\omega},
\end{equation}
and
\begin{equation}\label{2}
	\frac{\triangle Q}{\triangle E} = \frac{\mu}{\omega},
\end{equation}
where $\triangle E$ and $\triangle J$ correspond to the energy and angular momentum of the scalar field, and $\triangle Q$ defines the associated charge parameter. Assuming that the field is absorbed as well as scattered by the BH, and that the scattered part contributes no angular momentum or charge, we arrive at

\begin{equation}\label{3}
	\triangle \mathcal{M} < \triangle E, \qquad 
	\frac{\triangle Q}{\triangle E} = \frac{\mu}{\omega}, \qquad 
	\frac{\triangle J}{\triangle E} = \frac{m}{\omega}.
\end{equation}
Furthermore, using (\ref{3}) together with the conservation of charge and angular momentum, we obtain
\begin{equation}\label{4}
	E > \frac{a\triangle J + Q\triangle Qr_{+}}{a^{2} + r_{+}^{2}},
\end{equation}
alternatively, by substituting (\ref{1}) and (\ref{2}) into (\ref{4}), we get
\begin{equation}\label{5}
	\omega \geq \omega_\mathrm{Sl}=\frac{am}{a^{2} + r_{+}^{2}} 
	+ \frac{ Q\triangle Q r_{+}}{a^{2} + r_{+}^{2}},
\end{equation}
where $\omega_\mathrm{Sl}$ is called the superradiance limit.


	\subsection{overspinning extremal Kerr--Newman--MOG black hole}

		A KN--MOG--BH is extremal if it satisfies
 		\begin{equation}\label{deline}
			\triangle_\mathrm{In(E)}=\mathcal{M}^2-\sqrt{{\alpha}+1} \sqrt{J^2+\mathcal{M}^2 Q^2}=0.
		\end{equation}
	

		We investigate the perturbation of an extremal KN--MOG--BH by a scalar field to check the possibility of overspinning it into a naked singularity. The participation of the incoming wave to the angular momentum and the energy of the BH is given by (\ref{1}). For overspinning to take place, the parameters of the incoming 
		mode must be chosen so that $\triangle_\mathrm{Fin} < 0$ after the interaction. 
		More specifically, the requirement is
		\\
\begin{equation}\label{delfine}
	\triangle_\mathrm{Fin(E)} = \mathcal{M}^2_\mathrm{Fin} 
	- \sqrt{\alpha+1}\,\sqrt{J^2_\mathrm{Fin}+\mathcal{M}^2_\mathrm{Fin}Q^2_\mathrm{Fin}} < 0,
\end{equation}

\noindent where $\mathcal{M}_\mathrm{Fin}= (\mathcal{M}+\triangle \mathcal{M})$, 
$J_\mathrm{Fin}=(J+\triangle J)$, and $Q_\mathrm{Fin}= (Q+\triangle Q)$. 
Substituting these into (\ref{delfine}), we obtain
\begin{equation}\label{delfine1}
	\triangle_\mathrm{Fin(E)} 
	= (\mathcal{M}+\triangle \mathcal{M})^2 
	- \sqrt{\alpha+1}\,\sqrt{(J+\triangle J)^2 + (\mathcal{M}+\triangle \mathcal{M})^2 (Q+\triangle Q)^2} < 0,
\end{equation}

		where $\triangle \mathcal{M}$, $\triangle J$ and $\triangle Q$ denote the energy, angular momentum and charge of the test field respectively. The BH's ADM mass is influenced by the energy of the incident test field.
		In order to maintain the validity of the particle approximation, the energy of the field must be significantly less than the mass of the BH.  In order to justify the test field approximation, we select $\triangle E=\mathcal{M}\epsilon_1+Q\epsilon_2$ for the incoming charged field with 
		$0<\epsilon_1, \epsilon_2\ll1$. This choice allows the derivation of the maximum frequency of the incident wave, which may be employed to overspin an extremal KN--MOG--BH.
		Substituting $\triangle J=\frac{m}{\omega}\triangle E$, $\triangle Q=Q\epsilon_2$, $Q=\mathcal{M}\beta$ and using (\ref{deline}), the condition (\ref{delfine1}) can be written as
	\begin{equation}\label{MaxE}
		\omega < \omega_\mathrm{Max(E)} =
		\frac{m(\epsilon_1+ \beta \epsilon_2)\sqrt{1+\alpha}}{\mathcal{M}\Big[(1+\epsilon_1+\beta \epsilon_2)^2
			\sqrt{1-\dfrac{\beta^2(1+\alpha)(1+\epsilon_2)^2}{(1+\epsilon_1+\beta \epsilon_2)^2}}
			-\sqrt{1-\beta^2(1+\alpha)}\Big]}.
	\end{equation}

		 where $0<\beta<1$ and $0<\alpha <\frac{1-\beta ^2}{\beta ^2}$. An extremal KN--MOG--BH can be overspun into a naked singularity by a scalar field if its frequency is less than the maximum value found in (\ref{MaxE}). This condition, by itself, is insufficient to induce overspinning. For this purpose, it must also be required that the incident wave is absorbed by the BH; that is, its frequency must exceed the superradiance limit. Both conditions must be satisfied simultaneously for overspinning to take place. The superradiance limit $\omega_\mathrm{Sl(E)}$ and $\triangle_\mathrm{Fin(E)}$ for the extremal KN--MOG--BH are obtained by setting $r_{+} = 1+\alpha$, $a = \tfrac{J}{1+\alpha}$ and $M=1$ (without loss of generality) in (\ref{5}) and (\ref{delfine1})
		 
		\begin{equation}\label{SLE}
			\omega_\mathrm{Sl(E)} = 
			\frac{m\sqrt{1+\alpha}\,\sqrt{1-\beta^2(1+\alpha)} + \beta^2(1+\alpha)\mathcal{M}^2\epsilon_2}
			{\mathcal{M}\left[(2+\alpha)-\beta^2(1+\alpha)\right]},
		\end{equation}
		
		\begin{equation}\label{delfine2}
			\begin{aligned}
				\triangle_\mathrm{Fin(E)} = \;& (1+\epsilon_1+\beta \epsilon_2)^2  \\
				& - \sqrt{1+\alpha}\;
				\sqrt{(1+\epsilon_1+\beta \epsilon_2)^2 \beta^2 (1+\epsilon_2)^2 
					+ \left(\sqrt{\tfrac{1}{1+\alpha}-\beta^2}
					+ \tfrac{m(\epsilon_1+\beta^2 \epsilon_2)}{\omega \mathcal{M}}\right)^2} \\
				& < 0 \,.
			\end{aligned}
		\end{equation}

\begin{figure}[h!] 
		\centering
		\begin{subfigure}[t]{0.45\textwidth}
			\centering
			\includegraphics[width=\linewidth,height=4cm]{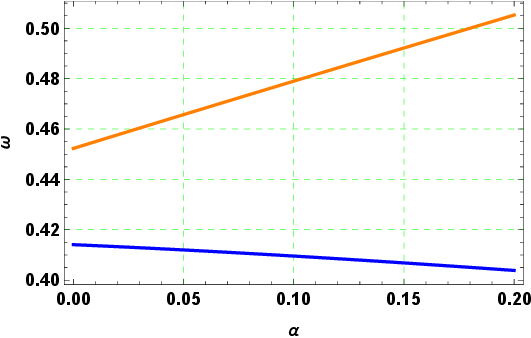}
			\caption{}
			\label{fig:A}
		\end{subfigure}
		\hfill
		\begin{subfigure}[t]{0.45\textwidth}
			\centering
			\includegraphics[width=\linewidth,height=4cm]{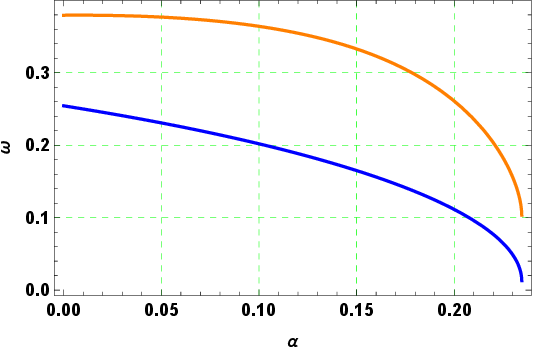}
			\caption{}
			\label{fig:E}
		\end{subfigure}
		\caption{$\omega_\mathrm{Max(E)}$ (orange) and $\omega_\mathrm{Sl(E)}$ (blue) for extremal BH when $\epsilon_1=\epsilon_2=0.01$, $m=1$, $\mathcal{M}=1.2$ and $\beta=0.5,0.9$ for (a)-(b) respectively.}
		\label{FigE1}
	\end{figure}

\begin{figure}[h!] 
		\centering
		\begin{subfigure}[t]{0.45\textwidth}
			\centering
			\includegraphics[width=\linewidth,height=4cm]{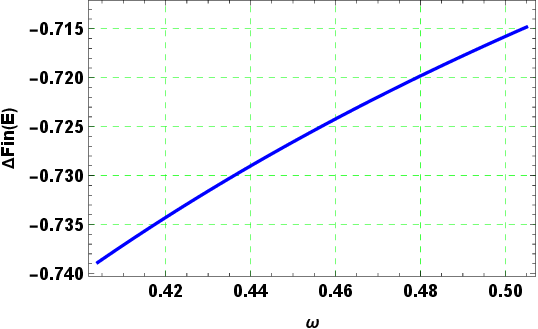}
			\caption{}
			\label{fig:A}
		\end{subfigure}
		\hfill
		\begin{subfigure}[t]{0.45\textwidth}
			\centering
			\includegraphics[width=\linewidth,height=4cm]{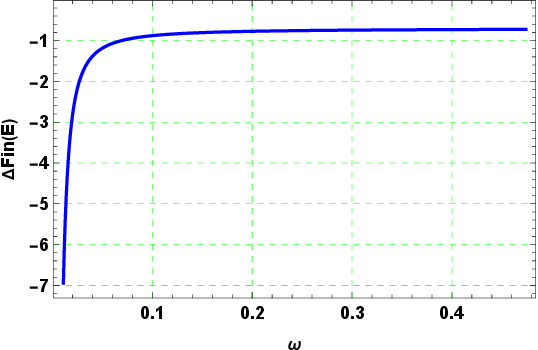}
			\caption{}
			\label{waqar}
		\end{subfigure}
		\caption{$\triangle_\mathrm{Fin(E)}$ for extremal BH for $\epsilon_1=\epsilon_2=0.01$, 
			$m=1$, $\mathcal{M}=1.2$, $\alpha=0.15, 0.05$, $\beta=0.5, 0.9$ 
			for (a)–(b), respectively.}
		\label{DelE}
	\end{figure}

		For overspinning to occur, the condition $\omega_\mathrm{Max(E)} > \omega_\mathrm{Sl(E)}$ must be satisfied, so that frequencies within the bound $(\omega_\mathrm{Sl(E)},\, \omega_\mathrm{Max(E)})$ can be employed to overspin an extremal KN--MOG--BH. We plot (\ref{MaxE}) and (\ref{SLE}) as functions of the MOG parameter $\alpha$, specifying $m, \mathcal{M}, \alpha, \beta, \epsilon_1$ and $\epsilon_2$ to see the manifestation of $\omega_\mathrm{Max(E)} > \omega_\mathrm{Sl(E)}$ (Fig.~\ref{FigE1}). We plot $\triangle_\mathrm{Fin(E)}$ as a function of the frequency $\omega$ 
		within the interval bounded by $\omega_\mathrm{Sl(E)}$ and $\omega_\mathrm{Max(E)}$, 
		by specifying $m, \mathcal{M}, \alpha, \beta, \epsilon_1,$ and $\epsilon_2$. 
		It is observed that $\triangle_\mathrm{Fin(E)} < 0$ throughout this admissible 
		frequency band,  which signals the violation 
		of the WCCC in this regime (Fig.~\ref{DelE}).
		
		\subsection{overspinning nearly extremal Kerr--Newman--MOG black hole}
		We now try to overspin a nearly extremal KN--MOG--BH by test scalar field. To parameterize the nearly extremal KN--MOG--BH, we define
		\begin{equation}\label{delinne}
			\frac{\sqrt{{\alpha}+1} \sqrt{J^2+\mathcal{M}^2 Q^2}}{\mathcal{M}^2}=(1-\gamma^2),
		\end{equation}
		where
		$\gamma = \sqrt{\epsilon \, \epsilon_{1}}, \; 0 < \epsilon_{1}, \epsilon_{2} \ll 1$. (\ref{delinne}) yields that
		\begin{equation}\label{waqar1}
			\triangle_\mathrm{In(NE)}=\sqrt{{1+\alpha}} \sqrt{J^2+\mathcal{M}^2 Q^2}=\mathcal{M}^2(1-\gamma^2).
		\end{equation}
A test field is sent from infinity, and we demand that $\triangle_{\mathrm{Fin(NE)}} < 0$ after the interaction, ensuring that the resulting spacetime parameters correspond to a naked singularity.
	\begin{equation}\label{waqar2}
		\triangle_{\mathrm{Fin(NE)}}
		= \sqrt{1+\alpha}\sqrt{(J+\triangle J)^2 + (\mathcal{M} + \triangle E)^2 (Q+\triangle Q)^2}
		- (1-\gamma^2)(\mathcal{M} + \triangle E)^2 <0.
	\end{equation}
		 By substituting 
	$\triangle J = \tfrac{m}{\omega}(\mathcal{M} \epsilon_{1} + Q \epsilon_{2}), \; 
	\triangle Q = Q \epsilon_{2}, \; Q = \mathcal{M} \beta$. Once again, we take the energy of the incoming wave to be $\triangle E = \mathcal{M}\epsilon_1 + Q\epsilon_2$ and using (\ref{waqar1}), the condition (\ref{waqar2}) for nearly extremal BH can be written as

\begin{align}\label{delfine2}
	\triangle_{\mathrm{Fin(NE)}} &= 
	(1+\epsilon_1+\beta \epsilon_2)^2 \nonumber\\
	&\quad - \sqrt{1+\alpha} \; 
	\Bigg[ 
	\sqrt{ 
		(1+\epsilon_1+\beta \epsilon_2)^2 \beta^2 (1+\epsilon_2)^2 
		+ \left( 
		\sqrt{\frac{(1-\gamma^2)^2}{1+\alpha} - \beta^2} 
		+ \frac{m(\epsilon_1+\beta^2 \epsilon_2)}{\omega \mathcal{M}} 
		\right)^2 
	} 
	\Bigg] < 0
\end{align}

		Following the same procedure as in the extremal case, one can directly obtain $\omega_\mathrm{Max(NE)}$ from (\ref{delfine2}) for a scalar field incident on a nearly extremal KN--MOG--BH, expressed in a parametrized form that can overspin the BH into a naked singularity
	\begin{align}\label{wmaxne}
		\omega_\mathrm{Max(NE)} &= 
		\frac{m \sqrt{1+\alpha} (\epsilon_1 + \beta \epsilon_2)}
		{\mathcal{M} \Bigg[
			(1+\epsilon_1+\beta \epsilon_2)^2 
			\sqrt{ 1 - \frac{\beta^2 (1+\alpha) (1+\epsilon_2)^2}{(1+\epsilon_1+\beta \epsilon_2)^2} } 
			- \sqrt{ (1-\gamma^2)^2 - \beta^2 (1+\alpha) } 
			\Bigg]}
	\end{align}
	As in the extremal case, condition (\ref{wmaxne}) alone is insufficient for overspinning. The frequency of the incoming wave must also exceed the superradiance limit. When $\omega_\mathrm{Sl(NE)} < \omega_\mathrm{Max(NE)}$, a range of frequencies $\omega_\mathrm{Sl(NE)} < \omega < \omega_\mathrm{Max(NE)}$ exists that can overspin a nearly extremal KN--MOG--BH. $\omega_\mathrm{Sl(NE)}$ is obtained in parameterized form
	
\begin{align}\label{wslne}
	\omega_\mathrm{Sl(NE)} &= 
	\frac{
		m \sqrt{1+\alpha} \, 
		\sqrt{(1-\gamma^2)^2 - \beta^2 (1+\alpha)}
		+ \beta^2 (1+\alpha) \mathcal{M}^2 \epsilon_2
		+ \gamma \beta^2 \epsilon_2 \mathcal{M} \sqrt{2-\gamma^2}
	}{
		\mathcal{M} \Big[ 
		(2+\alpha) + 2 \gamma \sqrt{1+\alpha} \, \sqrt{2-\gamma^2} - \beta^2 (1+\alpha) 
		\Big]
	}
\end{align}

\begin{figure}[h!] 
		\centering
		\begin{subfigure}[t]{0.45\textwidth}
			\centering
			\includegraphics[width=\linewidth,height=4cm]{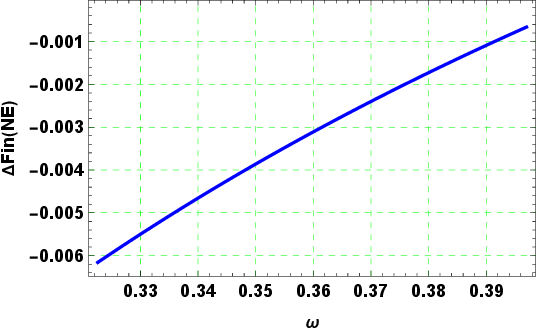}
			\caption{}
			\label{fig:A}
		\end{subfigure}
		\hfill
		\begin{subfigure}[t]{0.45\textwidth}
			\centering
			\includegraphics[width=\linewidth,height=4cm]{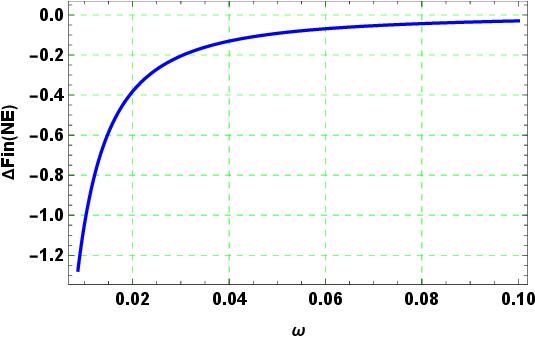}
			\caption{}
			\label{waqar}
		\end{subfigure}
		\caption{$\triangle_\mathrm{Fin(NE)}$ for nearly extremal BH for $\epsilon_1=\epsilon_2=0.01$, 
			$m=1$, $\mathcal{M}=1.2$, $\alpha=0.15, 0.20$, $\beta=0.5, 0.9$ 
			for (a)–(b), respectively.}
		\label{DelNE}
	\end{figure}

		We plot (\ref{wmaxne}) and (\ref{wslne}) as functions of the MOG parameter $\alpha$, specifying $m, \mathcal{M}, \alpha, \beta, \gamma, \epsilon_1$ and $\epsilon_2$ to see the manifestation of $\omega_\mathrm{Max(E)} > \omega_\mathrm{Sl(E)}$ (Fig.~\ref{FigE1}). It is observed that $\triangle_\mathrm{Fin(NE)} < 0$ throughout this admissible 
		frequency band,  which signals the violation 
		of the WCCC in this regime (Fig.~\ref{DelNE}).
	\begin{figure}[h!] 
		\centering
		\begin{subfigure}[t]{0.45\textwidth}
			\centering
			\includegraphics[width=\linewidth,height=4cm]{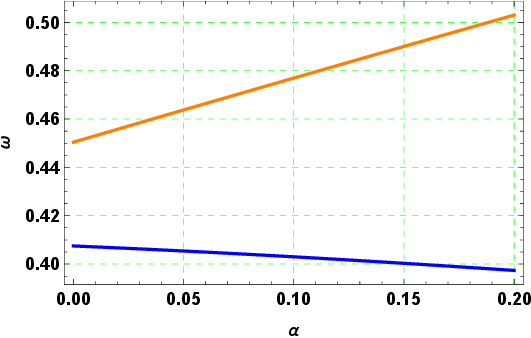}
			\caption{}
			\label{}
		\end{subfigure}
		\hfill
		\begin{subfigure}[t]{0.45\textwidth}
			\centering
			\includegraphics[width=\linewidth,height=4cm]{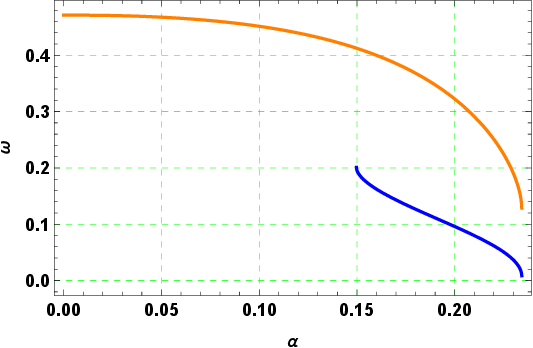}
			\caption{}
			\label{}
		\end{subfigure}
		\caption{$\omega_\mathrm{Max(NE)}$ (orange) and $\omega_\mathrm{Sl(NE)}$ (blue) for nearly extremal BH when $\epsilon_1=\epsilon_2=0.01$, $m=1$, $\mathcal{M}=1.2$ and $\beta=0.5,0.9$ for (a)-(b) respectively.}
		\label{FigNE1}
	\end{figure}

		 \section{Backreaction effects}\label{effects}
		 Will \cite{will1974perturbation}, in a foundational analysis, established that the infall of test particles towards a BH leads to a rise in the angular velocity of its horizon due to frame-dragging effects. The associated change in angular velocity is approximated by
		 \begin{equation}
		 	\triangle \omega  = \frac{\triangle J}{4 M^3}
		 \end{equation}
		 where $\triangle J$ represents the angular momentum carried by the test particle or field, while $M$ stands for the BH mass. It is important to note that the BH only gains angular momentum once the test particle or field has been absorbed, and not beforehand. If this were true, the overspinning of a nearly extremal BH would occur before absorption, since the spin parameter increases but the mass remains fixed. In reality, what changes prior to absorption is solely the angular velocity of the event horizon. Backreaction effects in scattering process arise as a consequence of the change in angular velocity. With the rise in the limiting frequency of superradiance, the absorption of field modes capable of overspinning the BH becomes forbidden.
		 
		\subsection{Backreaction effects for extremal Kerr--Newman--MOG black hole }
\begin{figure}[h!]
			\centering
			\includegraphics[width=0.45\textwidth]{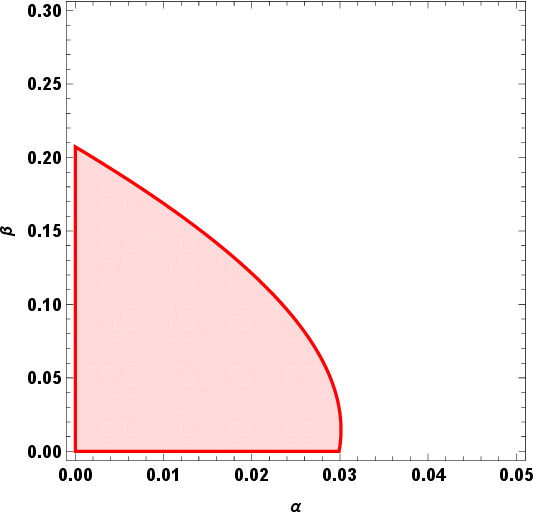}
			\caption{Region in the $(\alpha,\beta)$ plane where backreaction prevents overspinning the KN--MOG--BH for $m=1$, $\epsilon_1=\epsilon_2=0.01$.}
			\label{fig:regionPlot}
		\end{figure}

	 To evaluate the backreaction, we analyze an extremal BH perturbed by a test field whose frequency $\omega$ is taken to be arbitrarily close to, but just below, the $\omega_{\rm Max(E)}$. The test field will be absorbed by the BH since $\omega > \omega_{\rm Sl(E)}$, and its absorption may overspin the BH, leading to the formation of a naked singularity. As the test field moves closer to the BH, the event horizon’s angular velocity rises by an amount
	 
	 
	 \begin{equation} \label{BRE}
	 	\triangle\omega_\mathrm{(E)}=\frac{\triangle J}{4\mathcal{M}^3}=\frac{m(\epsilon_1+\beta \epsilon_{2})}{4\mathcal{M}^2\omega_\mathrm{Max(E)}},
	 \end{equation}
where we have used $\omega \approx \omega_{\rm Max(E)}, \; 
\triangle J = \tfrac{m}{\omega}\,\triangle E, \; 
\triangle E = \mathcal{M}\,\epsilon_{1} + Q\,\epsilon_{2}, \; 
Q = \mathcal{M}\,\beta.$ The critical frequency for superradiance is shifted upward by an amount $\triangle\omega_\mathrm{(E)}$. If the modified superradiant limit $\omega_{\rm Sl(E)} + \triangle\omega_\mathrm{(E)}
$ already exceeds the frequency of the incoming field at $\omega \approx \omega_{\rm Max(E)}$, then it will certainly surpass the field frequency throughout the interval $\omega_{\rm Sl(E)} < \omega < \omega_{\rm Max(E)}$, since the shift $\triangle\omega_\mathrm{(E)}$ becomes larger in that range. Hence, we evaluate the backreaction effects at $\omega \approx \omega_{\rm Max(E)}$ for certain $\alpha$ and $\beta$, imposing the condition
\begin{equation}\label{condition}
	\omega_{\rm Sl(E)} + \triangle\omega_\mathrm{(E)}\ge \omega,
\end{equation}
where $\omega \approx \omega_{\rm Max(E)}$, $\omega_{\rm Sl(E)}$ and $\triangle\omega_{\rm (E)}$ are given by equations (\ref{MaxE}), (\ref{SLE}) and (\ref{BRE}), respectively. We plot (\ref{condition}) to examine the region in the $(\alpha,\beta)$ plane where backreaction prevents overspinning of the extremal KN--MOG--BH for $\epsilon_1=\epsilon_2=0.01$ (Fig.\ref{fig:regionPlot}). We take $m = 1$ in (\ref{condition}), since these modes exhibit the maximum absorption probability, whereas the $m = 0$ modes, which carry no angular momentum to the BH, are omitted.
 There is no need to assign a value to $\mathcal{M}$, since it cancels out in the inequality. In contrast to the uncharged case analyzed by  Koray \cite{duztacs2020overspinning}, where the upper bound on the MOG parameter is 
 $\alpha \leq 0.0299$, our analysis shows that the inclusion of the charge parameter $\beta$ extends the admissible parameter 
 space to $0 \leq \alpha \leq 0.03$ with $0 \leq \beta \leq 0.21$. Hence, the presence of charge modifies the overspinning region 
 by broadening the allowed values of $\alpha$ while simultaneously constraining $\beta$.

	\subsection{Backreaction effects for nearly extremal Kerr--Newman--MOG black hole}\label{brenext}
	A similar approach can be applied to estimate the backreaction effects in the case of nearly extremal BH. Once again, it is required that the new superradiance limit remains greater than the frequency of the incident field, ensuring that overspinning does not occur. For nearly extremal BH, the modified superradiance limit is expressed as
	\begin{equation} \label{BRNE}
		\triangle\omega_\mathrm{(NE)}=\frac{\triangle J}{4\mathcal{M}^3}=\frac{m(\epsilon_1+\beta \epsilon_{2})}{4\mathcal{M}^2\omega_\mathrm{Max(NE)}},
	\end{equation}
	where we have used $\omega \approx \omega_{\rm Max(NE)}, \; 
	\triangle J = \tfrac{m}{\omega}\,\triangle E, \; 
	\triangle E = \mathcal{M}\,\epsilon_{1} + Q\,\epsilon_{2}, \; 
	Q = \mathcal{M}\,\beta.$
	Again we demand that
	\begin{equation}\label{condition1}
		\omega_{\rm Sl(NE)} + \triangle\omega_\mathrm{(NE)}\ge \omega,
	\end{equation}
	where $\omega \approx \omega_{\rm Max(NE)}$, $\omega_{\rm Sl(NE)}$ and $\triangle\omega_{\rm (NE)}$ are given by (\ref{wmaxne}), (\ref{wslne}) and (\ref{BRNE}), respectively. For near-extremal Kerr-MOG black holes with $\epsilon_1 = \epsilon_2 = 0.01$, $m = 1$, and $\gamma = 0.01$, the backreaction term $\triangle \omega_\mathrm{(NE)}$ is too small to satisfy the inequality (\ref{condition1}). Consequently, no feasible $\alpha$--$\beta$ region exists, and the backreaction is insufficient to prevent the BH from being overspun into a naked singularity. 
	
	\section{Conclusion}\label{conclusion}
We investigated the WCCC by analyzing the interaction of KN--MOG--BH with test field. Our analysis further utilizes the phenomenon of superradiance, which arises when scalar fields scatter off KN--MOG--BH. Superradiance is essential in a scattering process as it determines the lower limit for the frequency of a wave to ensure that it is absorbed by the black hole. Without this lower bound, modes with low energy and comparatively high angular momentum could be absorbed by the BH, thereby causing it to overspin. We first showed that test scalar fields with frequencies just above the superradiance limit can overspin both extremal and nearly extremal KN--MOG--BHs. Next, we incorporated the backreaction effects by considering the shift in the limiting frequency for superradiance. We have shown that the increase in the superradiance limit effectively prevents the overspinning of extremal KN--MOG--BHs for $m=1$ and $\epsilon_1=\epsilon_2=0.01$ (Fig.~\ref{fig:regionPlot}). We observed that the inclusion of charge extended the admissible range of $\alpha$ beyond the uncharged case discussed by Koray.
 However, for the same parameter values with $\gamma=0.01$, the backreaction fails to preserve the WCCC for nearly extremal KN--MOG--BHs.

\bibliographystyle{plainnat}
\bibliography{sn-bibliography}

\end{document}